\documentclass[aps,prd,twocolumn,superscriptaddress,showpacs,showkeys]{revtex4-2}
\usepackage{color,xcolor}
\usepackage{amssymb}
\usepackage{graphicx}
\usepackage{extarrows}

\newcommand{\bastar}{\begin{eqnarray*}}
\newcommand{\eastar}{\end{eqnarray*}}
\newskip\humongous \humongous=0pt plus 1000pt minus 1000pt

\newif\ifdtup

\relax
\newcommand{\W}{{\vec W}}
\newcommand{\n}{\hat n}

\newcommand{\hr}{\hat r}

\newcommand{\hD}{{\hat D}}

\newcommand{\bea}{\begin{eqnarray}}
\newcommand{\eea}{\end{eqnarray}}
\newcommand{\pd}{\partial}

\newcommand{\Int}{\displaystyle\int}
\newcommand{\A}{{\vec A}}
\newcommand{\hA}{{\hat A}}

\newcommand{\tA}{{\tilde A}}
\newcommand{\tC}{{\tilde C}}

\newcommand{\F}{{\vec F}}

\newcommand{\hF}{{\hat F}}

\newcommand{\mn}{{\mu\nu}}
\newcommand{\al}{\alpha}
\newcommand{\be}{\bar e}

\newcommand{\bal}{{\bar \alpha}}

\newcommand{\vsig}{{\vec \sigma}}
\newcommand{\vp}{{\varphi}}
\newcommand{\nn}{\nonumber}

\newcommand{\cL}{{\cal L}}

\newcommand{\cD}{{\cal D}}

\newcommand{\lam}{{\lambda}}

\begin{document}
\title{New Face of Standard Model: Neutral Magnetic Monopole}

\author{Y. M. Cho}
\email{ymcho0416@gmail.com}
\affiliation{School of Physics and Astronomy, Seoul National University, Seoul 08826, Korea}
\affiliation{Korea Academy of Science and Technology, Seungnam 13630, Korea}

\begin{abstract}
We show that the standard model has a totally new type of monopole which has the neutral magnetic charge $4\pi/\be$, where $\be$ is the neutral charge. This tells that the standard model has two types of monopoles, the well known Cho-Maison type which has the magnetic charge $4\pi/e$ and the neutral type which has the neutral magnetic charge $4\pi/\be$. This necessitates the new neutral charge quantization rule $\be=4\pi n/\bar g$, in addition to the Dirac's electric charge quantization rule $e=4\pi n/g$. The mass of the new monopole could be less than 
the Cho-Maison monopole, around 3.28 TeV. We show how to construct the neutral monopole and clarify the topological origin of the new monopole. We discuss the implications of the new monopoles in physics.
\end{abstract}
\keywords{Dirac monopole, Wu-Yang monopole, 'tHooft-Polyakov monopole, Cho-Maison monopole, neutral monopole, topology of 
the neutral monopole, neutral charge quantization rule, mass of the neutral monopole}
\maketitle

{\bf Introduction}---The standard model which unifies the electromagnetic and weak interactions has become one of the most successful theory in high energy physics \cite{wein}. With the recent discovery of Higgs particle at LHC, it has been widely asserted that the theory has passed the ``final" test \cite{LHC}. This has urged people to go ``beyond" the standard model. However, we believe that the standard model has not passed the final test, the topological test. 

By now it is well known that the standard model has the Cho-Maison monopole as the electroweak generalization of the Dirac monopole \cite{plb97,yang}. And this is within, not beyond, the standard model. 
This means that the true final test of 
the standard model should come from 
the confirmation of the topological objects of the theory, in particular the electroweak monopole.

After Dirac predicted the existence of 
the monopole, the monopole has become an obsession in physics \cite{dirac,cab}. But 
the monopole which should exist in the real world may not be the Dirac monopole but the electroweak monopole, because 
the Dirac monopole should change to the electroweak monopole in the course of 
the electroweak unification. This has triggered new studies on the electroweak monopole \cite{chen,epjc15,ellis,ak,bb,
mav,pta19,epjc20,gv,ms}. If detected, 
it will become the first magnetically charged topological elementary particle 
in the history of physics. For this 
reason MoEDAL and ATLAS at LHC and other 
experiments are actively searching for 
the monopole \cite{medal,atlas}.  

It has been known that the standard model has two conserved charges, the electric and neutral charges $e$ and $\be$. If so one might wonder if it could also have a monopole which carries the neutral magnetic charge $4\pi/\be$. This possibility of a neutral magnetic monopole has never been discussed in physics before.

{\it The purpose of this Letter is to show the exisitence of a totally new type of monopole which carries the neutral magnetic charge $4\pi/\be$ in the standard model. We argure that 
the monopole mass could be as small as 3.28 TeV, and discuss the topological origin of the neutral monopole. In doing so, we also show the existence 
of a new Cho-Maison monopole which 
has only the W boson dressing in 
the standard model.}  

In this connection it is interesting 
to notice that Nambu has proposed 
the existence of the neutral string 
known as the Z string in the standard 
model \cite{nambu,vacha,epjc24}, which could actually be viewed as the neutral monopole-antimonopole pair infinitely separated apart that we propose in this Letter. Indeed, the Z string strongly implies the existence of the neutral monopole in the standard model. But 
the neutral monopole in the standard model has never been discussed before. 

{\bf Two Types of Conserved Currents and Their Mixing in Standard Model}---To discuss the weak monopole we have to understand the existence of two types of conserved currents and their mixing in standard model, which provides the background reason for the existence of the neutral monopole. Consider the  Weinberg Lagrangian of the standard model \cite{wein},
\begin{gather}
{\cal L} =-|{\cal D}_\mu \phi|^2
-\frac{\lambda}{2}\big(|\phi|^2
-\frac{\mu^2}{\lambda}\big)^2
-\frac14\F_\mn^2
-\frac{1}{4}G_\mn^2, \nn \\
{\cal D}_\mu \phi =\big(\pd_\mu
-i\frac{g}{2} \vsig \cdot \A_\mu
-i\frac{g'}{2} B_\mu\big) \phi  \nn\\
=D_\mu \phi-i\frac{g'}{2} B_\mu \phi,
\label{lag0}
\end{gather}
where $\phi$ is the Higgs doublet, $\A_\mu$, $\F_\mn$ and $B_\mu$, $G_\mn$ are the gauge fields of SU(2) and hypercharge U(1), and $D_\mu$ is the covariant derivative of SU(2). Expressing $\phi$ by the Higgs scalar $\rho$ and unit Higgs doublet $\xi$
\begin{gather}
\phi = \dfrac{1}{\sqrt{2}} \rho~\xi,
~~~(\xi^\dagger \xi = 1),
\end{gather}
we have
\begin{gather}
\cL=-\frac{1}{2} (\pd_\mu \rho)^2
- \frac{\rho^2}{2} |{\cal D}_\mu \xi |^2
-\frac{\lam}{8}\big(\rho^2-\rho_0^2 \big)^2 \nn\\
-\frac14 \F_\mn^2 -\frac14 G_\mn^2,
\label{lag1}
\end{gather}
where $\rho_0=\sqrt{2\mu^2/\lam}$ is 
the vacuum expectation value of the Higgs field. Notice that the U(1) coupling 
of $\xi$ makes the theory a gauge theory 
of $CP^1$ field \cite{plb97}.

Choosing an arbitray direction $\n$ to be 
the Abelian direction in the SU(2) space, we have the Abelian decomposition of $\A_\mu$ to the restricted part $\hA_\mu$ and the valence part $\W_\mu$ which decomposes $\A_\mu$ to the non-topologcal Maxwellian $\tA_\mu$, topological Diracian $\tC_\mu$, and the gauge covariant 
$\W_\mu$ gauge independently \cite{prd80,
prl81,fadd,shab,zucc,kondo},
\begin{gather}
\A_\mu = \hA_\mu + \W_\mu,     \nn\\
\hA_\mu = \tA_\mu +\tC_\mu,
~~~\W_\mu =W^1_\mu ~\n_1 + W^2_\mu ~\n_2,  \nn\\
\tA_\mu=A_\mu \n,~~~\tC_\mu=-\frac{1}{g} \n\times \pd_\mu \n. 
\end{gather}
With this we have
\begin{gather}
\F_\mn=\hF_\mn + \hD _\mu \W_\nu - \hD_\nu
\W_\mu + g\W_\mu \times \W_\nu,   \nn\\
\hD_\mu=\pd_\mu+g \hA_\mu \times, \nn\\
\hF_\mn= \pd_\mu \hA_\nu-\pd_\nu \hA_\mu
+ g \hA_\mu \times \hA_\nu 
=F_\mn' \n, \nn \\
F'_\mn=F_\mn + H_\mn
= \pd_\mu A'_\nu-\pd_\nu A'_\mu,  \nn\\
F_\mn =\pd_\mu A_\nu-\pd_\nu A_\mu,  \nn\\
H_\mn = -\frac1g \n \cdot (\pd_\mu \n \times \pd_\nu \n)
=\pd_\mu C_\nu-\pd_\nu C_\mu,  \nn\\
A_\mu' = A_\mu+ C_\mu,~~~C_\mu \simeq -\frac1g \n_1 \cdot \pd_\mu \n_2.
\label{cdec}
\end{gather}
Moreover, with $\n=\xi^\dag \vsig \xi$ we have
\begin{gather}
C_\mu \simeq -\frac{2i}{g} \xi^\dagger \pd_\mu \xi,  \nn\\	
\cD_\mu \xi= \big[-i\frac{g}{2}(A'_\mu \n
+\W_\mu) \cdot \vsig 
-\frac{g'}{2 } B_\mu \big]~\xi, \nn\\
|\cD_\mu \xi|^2 =\frac{1}{4} (-gA'_\mu+g'B_\mu)_\mu^2
+\frac{g^2}{4} \W_\mu^2,
\label{id}
\end{gather}
so that we can remove the unit doublet $\xi$ from the Lagrangian (\ref{lag0})
completely and ``abelianize" it 
\begin{gather}
\cL = -\frac12 (\pd_\mu \rho)^2
-\frac{\lam}{8}\big(\rho^2-\rho_0^2 \big)^2 \nn\\
-\frac14 {F_\mn'}^2 -\frac14 {G_\mn}^2
-\frac12 \big|D_\mu' W_\nu 
-D_\nu' W_\mu \big|^2  \nn\\	
-\frac{\rho^2}{8} \big[(-gA_\mu'+g'B_\mu)^2 
+2 g^2 W_\mu^*W_\mu \big]  \nn\\
+i g F_\mn' W_\mu^* W_\nu 
+ \frac{g^2}{4}(W_\mu^* W_\nu 
-W_\nu^* W_\mu)^2,  \nn\\
D_\mu' =\pd_\mu +ig A_\mu',
~~~W_\mu =\frac{1}{\sqrt 2} (W^1_\mu +i W^2_\mu).
\label{lag2}
\end{gather}
From this we have the following equation of motion,
\begin{gather}
\pd_\mu^2\rho-\frac{\rho}{4}\Big((gA'_\mu
-g'B_\mu)^2+2g^2W_\mu^* W_\mu \Big)
=\frac{\lam}{2}(\rho^2-\rho_0^2) \rho, \nn\\
D_\mu'\Big(D_\mu'W_\nu -D_\nu'W_\mu \Big)
=\frac{g^2}{4}\rho^2 W_\nu  \nn\\
+igF_\mn' W_\mu +g^2 \Big(W_\mu^* W_\nu
-W_\nu^* W_\mu\Big) W_\mu,  \nn\\
\pd_\mu F_\mn' = j_\nu^{(w)},  \nn\\
\pd_\mu G_\mn = j_\nu^{(h)},
\label{wseq1}
\end{gather}
where $j_\nu^{(h)}$ and $j_\nu^{(w)}$ are the hypercharge and weak currents given by
\begin{gather}
j_\nu^{(w)} =ig\pd_\mu(W_\mu^* W_\nu
-W_\nu^* W_\mu) +\frac{g}{4}(gA_\nu'
-g'B_\nu) \rho^2  \nn\\
+ig\Big[W_\mu^* (D_\mu'W_\nu -D_\nu'W_\mu)
-(D_\mu'W_\nu -D_\nu'W_\mu)^* W_\mu \Big],  \nn\\
j_\nu^{(h)} =-\frac{g'}{4}(gA_\nu'
-g'B_\nu) \rho^2.
\label{j1}
\end{gather}
Notice that the last two equations of (\ref{wseq1}) tells that the theory has 
two conserved currents, the weak current $j_\mu^{(w)}$ of $A_\mu'$ and the hypercharge current $j_\mu^{(h)}$ of $B_\mu$.

Now, we can define the electromagnetic potential $A_\mu^{\rm (em)}$ and the neutral potential $Z_\mu$ by
\begin{gather}
\left(\begin{array}{cc} A_\mu^{\rm (em)} \\  Z_{\mu}
\end{array} \right)
=\frac{1}{\sqrt{g^2 + g'^2}} \left(\begin{array}{cc} g & g' \\
-g' & g \end{array} \right)
\left(\begin{array}{cc} B_{\mu} \\ A'_{\mu}
\end{array} \right)  \nn\\
= \left(\begin{array}{cc}
\cos\theta_{\rm w} & \sin\theta_{\rm w} \\
-\sin\theta_{\rm w} & \cos\theta_{\rm w}
\end{array} \right)
\left( \begin{array}{cc} B_{\mu} \\ A_\mu'
\end{array} \right),
\label{mix}
\end{gather}
and can express the Lagrangian (\ref{lag2}) in terms of the physical fields by
\begin{gather}
\cL= -\frac12 (\pd_\mu \rho)^2
-\frac{\lam}{8}\big(\rho^2-\rho_0^2 \big)^2
-\frac14 {F_\mn^{\rm (em)}}^2   \nn\\
-\frac12 \big|(D_\mu^{\rm (em)} 
+i\be Z_\mu) W_\nu
-(D_\nu^{\rm (em)} +i\be Z_\nu) W_\mu \big|^2  \nn\\
-\frac14 Z_\mn^2-\frac{\rho^2}{4} \big(g^2 W_\mu^*W_\mu
+\frac{g^2+g'^2}{2} Z_\mu^2 \big)   \nn\\
+i (e F_\mn^{\rm (em)}
+ \be Z_\mn) W_\mu^* W_\nu  
+ \frac{g^2}{4}(W_\mu^* W_\nu - W_\nu^* W_\mu)^2,  \nn\\
D_\mu^{\rm (em)}=\pd_\mu+ieA_\mu^{\rm (em)},
\label{lag3}
\end{gather}
where $e$ and $\be$ are the electric and neutral charges, 
\begin{gather}
e=\frac{gg'}{\sqrt{g^2+g'^2}},
~~~\be =\frac{g^2}{\sqrt{g^2+g'^2}}.
\end{gather}
Notice that $e/\be=g'/g$. Experimentally 
we have $\be \simeq 1.83~e$, so that 
the weak coupling is stronger than the electromagnetic coupling. Moreover, this confirms that the W boson has both electric and weak charges, so that it is doubly charged. We emphasize that (\ref{lag3}) 
is not the standard model in the unitary gauge. This is a gauge independent expression of the standard model. 

In the Lagrangian the W and Z bosons acquire the mass by the Higgs mechanism.
But here we have the mass generation without any symmetry breaking, spontaneous or not \cite{pta19,epjc20,epjc24}. To see this notice that (\ref{lag3}) is mathematically identical to (\ref{lag0}), so that it retains the full SU(2) and U(1) gauge symmetry of (\ref{lag0}). But the W and Z bosons acquire mass from the non-vanishing vacuum value of the Higgs scalar $\rho_0$, which does not break any symmetry. This confirms that the Higgs mechanism does not need any spontaneous symmetry breaking. 

From (\ref{lag3}) we obtain the following equations of motion
\begin{gather}
\pd^2 \rho-\big(\frac{e^2+\be^2}{2} W_\mu^*W_\mu
+\frac{e^2+\be^2}{4 \be^2}Z_\mu^2 \big)~\rho
=\frac{\lambda}{2}\big (\rho^2 -\rho_0^2 \big)~\rho,   \nn\\
\big(D_\mu^{\rm (em)}+i\be Z_\mu\big) 
\big[(D_\mu^{\rm (em)}+i\be Z_\mu) W_\nu \nn\\
-(D_\mu^{\rm (em)}+i\be Z_\mu) W_\mu \big] 
=\frac{(e^2+\be^2)}{4}\rho^2 W_\nu  \nn\\
+ i  W_\mu \big(e F^{\rm(em)}_\mn 
+ \be Z_\mn \big)
+ (e^2+\be^2) W_\mu(W_\mu^* W_\nu-W_\nu^* W_\mu),\nn\\
\pd_\mu F_\mn^{\rm(em)} = J_\mu^{(e)},  \nn\\
\pd_\mu Z_\mn =J_\nu^{(n)},
\label{wseq2}
\end{gather}
where $J_\mu^{(e)}$ and $J_\nu^{(n)}$ are 
the electromagnetic and neutral currents 
given by
\begin{gather}
J_\mu^{(e)} = ie \Big\{\pd_\mu 
(W_\mu^* W_\nu-W_\nu^* W_\mu)  \nn\\
+W_\mu^* \big[(D_\mu^{\rm (em)} +i\be Z_\mu) W_\nu 
-(D_\nu^{\rm (em)} +i\be Z_\nu) W_\mu \big] \nn\\
-W_\mu \big[(D_\mu^{\rm (em)} +i\be Z_\mu) W_\nu -(D_\nu^{\rm (em)} +i\be Z_\nu) W_\mu \big]^*  \Big\},\nn\\
J_\mu^{(n)} =i\be \Big\{\pd_\mu  (W_\mu^* W_\nu -W_\mu W_\nu^*)  \nn\\
+ W_\mu^* \big[(D_\mu^{\rm (em)} +i\be Z_\mu) W_\nu 
-(D_\nu^{\rm (em)} +i\be Z_\nu) W_\mu \big] \nn\\
-W_\mu \big[(D_\mu^{\rm (em)} +i\be Z_\mu) W_\nu -(D_\nu^{\rm (em)} 
+i\be Z_\nu) W_\mu \big]^*  \Big\} \nn\\
+\frac{(e^2+\be^2)^2}{4 \be^2} \rho^2 Z_\nu.
\label{j2}
\end{gather}
This should be compared with the equations 
of motion (\ref{wseq1}) obtained from (\ref{lag2}). The two sets of currents in (\ref{j1}) and (\ref{j2}) are related 
by
\begin{gather}
\left(\begin{array}{cc} J_\mu^{\rm (e)} \\  J_\mu^{(n)} \end{array} \right)
= \left(\begin{array}{cc}
\cos\theta_{\rm w} & \sin\theta_{\rm w} \\
-\sin\theta_{\rm w} & \cos\theta_{\rm w}
\end{array} \right)
\left(\begin{array}{cc} j_\mu^{(h)} \\ 
j_\mu^{(w)} \end{array} \right).
\label{jmix}
\end{gather}
The existence of two sets of currents and their mixing in the standard model has not been properly appreciated so far. But this will have deep implication in the following when we discuss the new monopole in the standard model.

{\bf Cho-Maison Monopole: A New Result}---Before we discuss the neutral monopole we update the well known Cho-Maison monopole in the standard model to add a new Cho-Maison monopole solution which has not been discussed before. 

Consider the following Cho-Maison monopole ansatz in the spherical coordinates $(t,r,\theta,\varphi)$ \cite{plb97,yang}
\begin{gather}
\phi=\frac{1}{\sqrt 2} \rho(r)~\xi,
~~~\xi= i \left(\begin{array}{cc}
\sin \dfrac{\theta}{2}~\exp(-i\varphi) \\
- \cos \dfrac{\theta}{2} \end{array} \right), \nn\\
\hA_\mu =-\frac{1}{g}~\hr \times \pd_\mu \hr,
~~~\W_\mu =\frac{1}{g} f(r)~\hr \times \pd_\mu \hr,   \nn\\
B_{\mu} =-\frac{1}{g'}(1-\cos\theta) \pd_\mu \varphi.
\label{ans0}
\end{gather}
Notice that $\A_\mu =\hA_\mu+\W_\mu$ 
is exactly the 'tHooft-Polyakov monopole ansatz and $B_\mu$ describes the Dirac monopole \cite{wu,prl80}. 
In terms of physical fields the ansatz becomes
\begin{gather}
\rho =\rho(r),  
~~~W_\mu= \frac{i}{g} \frac{f}{\sqrt 2}e^{i\varphi} (\pd_\mu \theta 
+i \sin\theta \pd_\mu \varphi), \nn\\
A_\mu^{\rm (em)}=-\frac1e (1-\cos\theta)\pd_\mu \varphi,
~~~Z_\mu= 0.
\label{ans1}
\end{gather}
This clearly shows that the ansatz is for the electromagnetic monopole.

With the ansatz we have the following equations of motion
\begin{gather}
\ddot \rho+\frac{2}{r}\dot \rho -\frac{f^2}{2r^2} \rho
=\frac{\lambda}{2} \big(\rho^2
-\rho_0^2 \big) \rho, \nn \\
\ddot f -\frac{f^2 -1}{r^2}~f
=\frac{g^2}{4} \rho^2~f,
\label{cmeq}
\end{gather}
which has a singular solution
\begin{gather}
f=0,~~~\rho=\rho_0 =\sqrt{2\mu^2/\lambda},  \nn\\
A_\mu^{\rm (em)} = -\frac{1}{e}(1-\cos \theta)
\pd_\mu \varphi,~~~Z_\mu=0.
\label{cmon}
\end{gather}
This is the naked Cho-Maison monopole in the standard model which has the magnetic charge $4\pi/e$ (not $2\pi/e$), which could be identified as a Schwinger type monopole. 

\begin{figure}
\includegraphics[height=4.5cm, width=8cm]{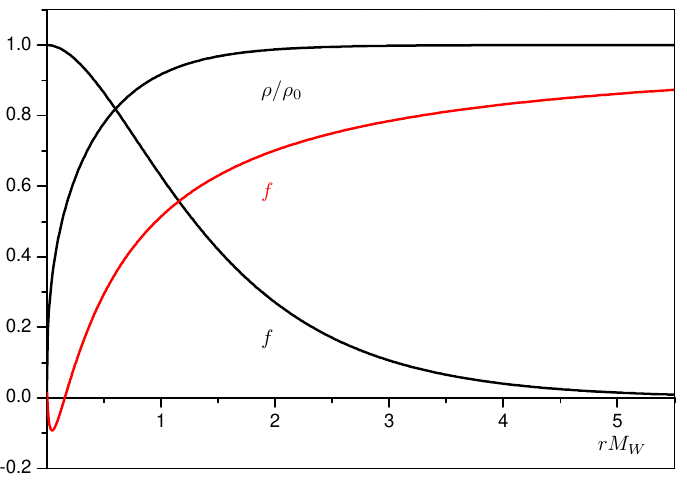}
\caption{\label{ncmm} The neutral monopole solution which has the magnetic charge $4\pi/\be$ in the standard model. The W boson dressing of the monopole is shown in red curve, which also describes the W boson dressing of the new Cho-Maison monopole. For comparison we plot the well known Cho-Maison monopole solution which has the Higgs and W boson dressing in black curves.}
\end{figure}

With the boundary condition
\begin{gather}
\rho(0)=0,~~\rho(\infty)=\rho_0,  \nn\\
f(0)=1,~~f(\infty)=0,
\label{bc0}
\end{gather}
we can integrate (\ref{cmeq}) and have
the absolutely stable Cho-Maison monopole solution carrying the magnetic charge $q_m=4\pi/e$ which has 
a non-trivial Higgs and W boson 
dressing \cite{plb97,epjc15}. This is shown in Fig. \ref{ncmm} in black curves. Moreover, we can also construct the anti-monopole solution, and generalize it to a dyon and anti-dyon solutions which have an extra electric charge \cite{epjc15,pta19}. 

The monopole has the following energy \cite{plb97,epjc15,pta19},
\begin{gather}
E =E_0 +E_\rho +E_W +E_{\rho W},  \nn\\	
E_0 = \frac{4\pi}{{g'}^2} M_W \Int_0^\infty ~\frac{1}{2 r^2}\frac{dr}{M_W},  \nn\\
E_\rho =M \Int_0^\infty 2 r^2
\Big[\big(\frac{\dot \rho}{\rho_0} \big)^2 +\frac{\lam \rho_0^2}{4} \big((\frac{\rho}{\rho_0})^2 
-1 \big)^2 \Big] M_W dr, \nn\\
E_W= M \Int_0^\infty 
\Big[\dot f^2 +\frac{(f^2-1)^2}{2r^2} \Big] \frac{dr}{M_W}, \nn\\
E_{\rho W} = M \Int_0^\infty f^2 (\frac{\rho}{\rho_0})^2 M_W dr,
\label{cme}
\end{gather}
where $M=(4\pi/g^2) M_W\simeq 2.6~{\rm TeV}$. The boundary condition (\ref{bc0}) makes $E_\rho$, $E_W$, and $E_{\rho W}$ finite. But $E_0$ becomes infinite, so that it has infinite energy. This is because $B_\mu$ in (\ref{ans0}) has the monopole point singularity at the origin. This means that we can not estimate the monopole mass classically. However, notice that the energy scale is fixed by $M$.  

There have been many efforts to estimate the monopole mass. Zeldovich and Khlopov first predicted the mass 
to be around 11 TeV, $1/\al$ times 
the W boson mass, arguing that the monopole mass comes from the same Higgs mechanism which makes the W boson massive, except that here the coupling is $4\pi/e$ \cite{zel}. After this there have been other estimates, notably based on the scaling argument and the charge screening mechanism, 
and the general consensus now is that the mass could be around 4 to 11 
TeV \cite{epjc15,pta19,ellis,bb,zel}.

Before we leave this section we show that there is another interesting Cho-Maison type monopole in the standard model. To do this, notice that in the absence of $\phi$ (i.e., when $\rho=0$), the Cho-Maison monopole equation (\ref{cmeq}) reduces to
\begin{gather}
\ddot f -\frac{f^2 -1}{r^2}~f =0.
\label{ncmeq}
\end{gather}
We can integrate this with the boundary 
condition
\begin{gather}
f(0) =0,~~~f(\infty) =1,
\label{ncmbc}
\end{gather}
and obtain a new Cho-Maison monopole solution which has only the W boson dressing, which carries the magnetic charge $4\pi/e$. This is shown in Fig. \ref{ncmm} in red curve. Notice that the boundary condition (\ref{ncmbc}) is different from the boundary condition (\ref{ncmbc}) in (\ref{bc0}).

The energy of this monopole is given by
\begin{gather}
E =E_{CM} +E_W,  \nn\\	
E_{CM} =\frac{4\pi}{{g'}^2} M_W \Int_0^\infty ~\frac{1}{2 r^2}\frac{dr}{M_W},  \nn\\
E_W =\frac{4\pi}{g^2} M_W \Int_0^\infty 
\Big[\dot f^2 +\frac{(f^2-1)^2}{2r^2} \Big] \frac{dr}{M_W},
\label{ncme}
\end{gather}
where $E_0$ is the energy of the naked Cho-Maison monopole and $E_W$ is the energy of the W boson dressing. But unlike (\ref{cme}) both $E_0$ and $E_W$ become divergent here, because the boundary condition (\ref{ncmbc}) makes $E_W$ divergent near the origin.

This tells that there are actually three Cho-Maison type monopole solutions in the standard model, the naked one, the one which has the W and Higgs boson dressing, and the one which has only thw W boson dressing. 

{\bf Neutral Monopole}---Now we are ready 
to discuss a totally different type of new monopole solution in the standard model. 
To show this, we choose the ansatz
\begin{gather}
\phi =0,   
~~~\hA_\mu =-\frac1g~\hr \times \pd_\mu \hr, 
~~~\W_\mu =\frac{f(r)}{g} ~\hr \times \pd_\mu \hr,  \nn\\
B_\mu =\frac{g'}{g^2} (1-\cos \theta)
\pd_\mu \vp.
\label{nans0}
\end{gather}
Notice that $\phi=0$ means that the anzstz has no Higgs field. In terms of the physical fields, this is expressed by
\begin{gather}
\rho =0,~~~W_\mu= \frac{i}{g} \frac{f}{\sqrt 2} (\pd_\mu \theta +i \sin\theta \pd_\mu \vp), \nn\\
A_\mu^{\rm (em)} =0,
~~~Z_\mu=-\frac{1}{\be}(1-\cos \theta)
\pd_\mu \vp,
\label{nans1}
\end{gather}
This should be compared with the Cho-Maison monopole ansatz (\ref{ans1}). The difference is that here $Z_\mu$, in stead of $A_\mu^{\rm (em)}$, has the monopole. This clearly tells that the ansatz is for the neutral magnetic monopole. 
	
With the ansatz we have the following 
equation,
\begin{gather}
\ddot f -\frac{f^2 -1}{r^2}~f =0,
\label{ceq}
\end{gather}
which is identical to (\ref{ncmeq}). This has the trivial solution $f=0$, which describes the naked neutral monopole,
\begin{gather}
\rho =0,~~~W_\mu=0,  \nn\\
A_\mu^{\rm (em)} =0,
~~~Z_\mu=-\frac{1}{\be}(1-\cos \theta)
\pd_\mu \vp.
\label{nans1}
\end{gather}
We can also integrate this with the boundary condition (\ref{ncmbc}), and obtain the neutral magnetic monopole solution which carries the neutral magnetic charge $4\pi/\be$. This is shown in Fig. \ref{ncmm} in red curve. Notice that, although the W boson dressing is identical to the W boson dressing of Cho-Maison monopole that 
we discussed above, physically this solution describes a totally different monopole, the neutral monopole in the standard model.

Near the origin the solution has the following behavior \cite{yang1},
\begin{gather}
f(r)\simeq r^{1/2}\Big(A\cos\big(\frac{\sqrt{3}}{2}
\ln r \big) +B\sin\big(\frac{\sqrt{3}}{2}\ln r \big) \Big)\nn\\
+O\left(r^{1/2}\right),
\end{gather}
where $A$ and $B$ are the integration constants. But asymptotically it behaves as
\begin{gather}
f(r)\simeq 1
-\frac{C}{r}+O\Big(\frac{1}{r^2} \Big),
\end{gather}
where $C$ is another integration constant. This tells that, unlike the well known Cho-Maison monopole which has the Higgs and W boson dressing, 
the size of the the neutral monopole 
is not fixed by the W boson mass.

One might wonder what is the topological origin of the neutral monopole. It comes from the $\pi_1(S^1)$ monopole topology of 
the neutral U(1). This must be clear 
from the ansatz (\ref{nans1}), where 
$Z_\mu$ is given by the Dirac monopole potential which has the magnetic charge $4\pi/\be$. In comparison, the Cho-Maison monopole has the $\pi_1(S^1)$ monopole topology of the electromagnetic U(1). This is clear from (\ref{ans1}). This tells that 
the new monopole and the Cho-Maison monopole have different monopole topology.   

The energy of the neutral monopole 
is given by
\begin{gather}
E =E_N +E_W,  \nn\\	
E_N =\frac{4\pi}{\bar e^2} M_W \Int_0^\infty ~\frac{1}{2 r^2}\frac{dr}{M_W},  \nn\\
E_W =\frac{4\pi}{g^2} M_W \Int_0^\infty 
\Big[\dot f^2 +\frac{(f^2-1)^2}{2r^2} \Big] \frac{dr}{M_W},
\label{nme}
\end{gather}
where $E_N$ is the energy of the naked neutral monopole and $E_W$ is the energy of the W boson dressing. Both $E_N$ and $E_W$ are divergent at the origin. So, just like the Cho-Maison monopole, the neutral monopole has an infinite energy classically. Notice that this energy is different from the energy (\ref{ncme}) of the Cho-Maison monopole which has 
the same W boson dressing. 

\begin{figure}
\includegraphics[height=4.5cm, width=8cm]{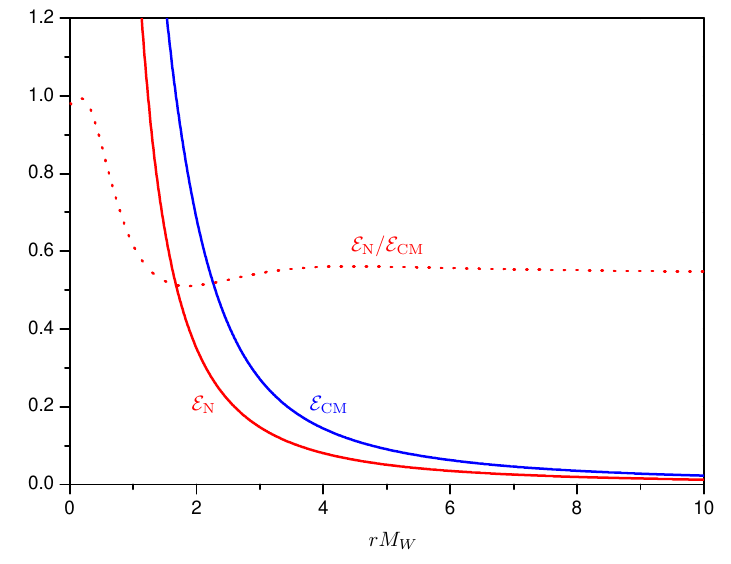}
\caption{\label{enden} The energy density of the electroweak monopoles. The blue and red curves represent the energy densities of the Cho-Maison monopole which has the Higgs and W boson dressing and the neutral monopole which has the W boson dressing, and 
the dotted line represents the ratio between the two energy densities.}
\end{figure}

This tells that the standard model has two types of monopoles, the Cho-Maison type electromagnetic monopoles which carry the magnetic charge $4\pi/e$ and the neutral type monopoles which carry the magnetic charge $4\pi/\be$. There are three Cho-Maison type monopoles, 
the naked one, the one dressed by 
the Higgs and W bosons, and the one dressed only by the W boson. And there are two neutral type monopoles, the naked one and the one with the W boson dressing.  

Is there any way to estimate the mass of the neutral monopole? Just as we could estimate the mass of the Cho-Maison monopole, there are various ways to do that. For example, we could also argue that the mass of the neutral monopole could be $1/\bal$ times the W boson mass, where $\bal=\be^2/4\pi$ is the fine structure constant of the neutral charge. With $\be \simeq 1.83 \times e$, we have $\bal \simeq 3.35 \times \al$. This implies that the neutral monopole mass could be much smaller than the mass of the Cho-Maison monopole, around 3.28 TeV. 
	
Admittedly one could question the reliability of this estimate. But the important point here is that the mass of the neutral monopole could be similar to, or even smaller than, the mass of the Cho-Maison monopole. This is backed up by Fig. \ref{enden}, which shows that energy density density of the neutral monopole is everywhere less than 
that of the Cho-Maison monopole. 
 
{\bf Neutral Monopole and Nambu String}---We have shown the existence of the neutral monopole solving the equation of motion. Now, we show how to construct the neutral monopole from
Nambu's neutral string. It is well known that Dirac monopole originates from the Dirac string, the infinitely thin string carrying the quantized magnetic flux. When we cut it, the monopole flux appears at the cutted edge. But this is not a monopole but a magnetic flux coming out of the string. However, Dirac showed us how to make this string invisible with the Dirac's charge quantization condition. 

\begin{figure}
\includegraphics[height=6.5cm, width=4.5cm]{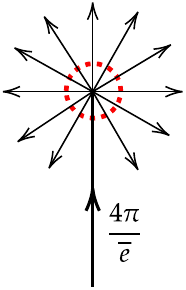}
\caption{\label{namo} The ``Nambu monopole" denoted by the dotted red circle which has the Z string tail carrying the neutral magnetic flux $4\pi/\bar e$ along the negative $z$-axis. We can obtain the neutral monopole which has no neutral magnetic flux tail, with the neutral charge quantization rule.}
\end{figure}

Remarkably, Nambu told us that the standard model has the Z string, a neutral string carrying the quantized neutral magnetic flux \cite{nambu,vacha,
epjc24}. Cutting this neutral string, we have the neutral monopole flux coming out of the neutral string. This is called the ``Nambu monopole" in the literature \cite{kib,eto}. But this is not a monopole, but a neutral magnetic flux originating from the string. This is shown in Fig. \ref{namo}. 

Notice, however, we could also remove this string imposing the neutral charge quantization condition as Dirac did. With this we obtain the neutral monopole with no string attached at all. This is the neutral monopole that we have discussed. 

Of course, the Nambu's Z string is not 
a singular string. But we can show that the standard model has a singular Z string, which makess the above logic for the existence of the neutral monopole valid. But this tells that we have two charge quantization rules in the standard model, the neutral charge quantization and the electric charge quantization. 

So, we can actually construct the neutral monopole from the neutral string, with the neutral charge quantization condition. This provides us another reason why the neutral monopole must exist in the standard model.

{\bf Discussion}---In this work we have shown the existance of the electrically neutral  monopole solutions in the standard model which has the neutral magnetic charge $4\pi/\be$. This 
tells that the standard model actually has two types of monopoles, the electromagnetic Cho-Maison monopole and the non-electromagnetic neutral monopole discussed in this paper. 

The existence of the two monopoles
in the standard model stems from the fact that the standard model has two Abelian subgroups, the U(1) subgroup of SU(2) and the hypercharge U(1), or equivalently the electromagnetic U(1) and the neutral U(1). This guarantees the existence of two conserved charges,
the electric and neutral charges, in the standard model. This strongly implies the existence of two types of monopoles. However, this provides only a theoretical possibility for two types of monopoles. The fact that the theory actually accommodates two types monopoles is unexpected. 

The electroweak monopoles could play important roles in physics. Amply produced during the electroweak phase transition they could turn to the primordial magnetic blackholes by the gravitational collapse, and play the fundamental role in the formation of the galaxies and large scale structures of the universe. Moreover, 
some of them (heavier ones) become the seed of the supermassive blackholes, and the lighter blackholes could even account for the dark matter of the universe \cite{arx24}. 

They could also play important roles 
in low energy physics. This is because the standard model could be viewed as the non-Abelian Ginzburg-Landau 
theory of two-gap ferromagnetic superconductivity \cite{pla23,ap24,
arx251}, since the Higge doublet could be interpreted to describe the spin doublet Cooper pair in two gap ferromagnetic superconductors. 
Moreover, the standard model could 
also describe the magnon electron spintronics \cite{arx251,arx252,
arx253}. This is because the Higgs doublet could also be interpreted as the charged spinon which describes the electron in electron spintronics. So, with the hypercharge U(1) as the electromagnetic U(1) and the non-Abelian SU(2) interaction as the magnon gauge interaction acting on the electron, we could view the Weinberg Lagrangian as the Lagrangian for the electron magnon spintronics. In this case the monopoles in the standard model could play important roles as 
the topological objects in these non-Abelian condensed matters.

An important issue on the neutral monopole in the standard model is 
how to estimate the mass of the neutral monopole. Our analysis suggests the mass of the neutral monopole could be 3.28 TeV, but definitely we need a more reliable estimate of the mass.

From the practical point of view, however, a most important problem 
is how can we confirm the existence of two types of monopoles in the standard model experimentally. As we have mentioned, the standard model has yet to pass the final test, the topological test. And so far it has generally been believed that this topological test could come from the experimental confirmation of the Cho-Maison monopole. The problem with this is that it is not clear that the present 14 TeV LHC could actually produce it or not, since the mass has been predicted 
to be 4 to 11 TeV \cite{epjc15,ellis,
ak,bb,mav,pta19,epjc20}. 

Now, the advent of the neutral 
monopole could change this situation drastically. First of all, with 
the neutral monopole we have more possibility for the topological test 
of the standard model. More importantly, if the mass of the neutral monopole becomes around 3 to 5 TeV, the present 14 TeV LHC energetically could produce this monopole in pairs. In this case ATLAS and CMS could in principle detect it, if the production probability is big enough for them to detect the monopole. 
If so, the first topological test of the standard model could come from the detection of the neutral monopole, not the Cho-Maison monopole. This provides a new motivation for ATLAS and CMS to look for the new monopole more seriously. 

Before we close, we emphasize that 
the neutral monopole is not just an isolated solution in the standard model. 
Mathematically exactly the same 
monopole solutions appear in QCD and  Georgi-Glashow model. It is well known that QCD has the Wu-Yang monopole. We can generalize this Wu-Yang monopole to have the chromon dressing \cite{cho}. And this chromonic monopole in QCD becomes the neutral monopole when embedded in the standard model. This tells that the dressed Wu-Yang monopole is the proto-type monopole of the new monopoles which appear in these non-Abelian gauge theories.

Our result implies that the standard model may have new territory which 
has yet to be explored. The details of the neutral monopole in the standard model and its relation to the new monopoles in QCD and Georgi-Glashow model will be discussed in a separate paper \cite{cho}.

{\bf Acknowledgments}

~~~This work is supported in part by Korea Academy of Science and Technology.

\end{document}